%% file: article.tex
\documentclass[12pt]{article}
\usepackage{epsfig}
\usepackage{lineno}

\input econfmacros.tex
\def\Title#1{\begin{center} {\Large {\bf #1} } \end{center}}

\begin{document}
%\linenumbers

\Title{Measurements of the top quark charge asymmetry at the LHC}

\bigskip\bigskip

%+\addtocontents{toc}{{\it D. Reggiano}}
%+\label{ReggianoStart}

\begin{raggedright}  

{\it Umberto De Sanctis, on behalf of the ATLAS and CMS collaborations\index{De Sanctis, U.}\\
Universit\`a degli studi di Udine \& INFN Gruppo Collegato di Udine}
\bigskip\bigskip
\end{raggedright}

\section{Introduction}

Since its discovery in 1995, the top quark is playing a key role in the understanding of Quantum Chromodynamics (QCD)  processes at high energies. The significant top quark pair production cross-section at the LHC  allows to deeply explore the production mechanisms and search for signals of New Physics processes beyond the Standard Model (SM). In this article, the top quark charge asymmetry measurements performed by the ATLAS~\cite{ATLAS_GEN} and CMS~\cite{CMS_GEN} experiments are presented. Results in single lepton and dilepton top decay channels for $pp$ collisions at 7 TeV center-of-mass energy using data collected in 2011 are shown.

\section{The top quark charge asymmetry}

At the LHC collider, the top quark pairs are produced mainly through gluon-gluon fusion. Only around 20\% of the events are produced from $q\bar{q}$ hard collisions, while the fraction coming from $qg$ partonic processes is almost negligible. The charge asymmetry $A_C$ is a manifestation of the forward-backward asymmetry when the CP invariance holds. It is a tiny NLO QCD effect ($A_C^{SM}$ = 0.0115$\pm$0.0006~\cite{THEOR}) present only in asymmetric initial states, like $q\bar{q}$ and $qg$. In the $t\bar{t}$-system center-of-mass frame, the effect of the charge asymmetry is that tops (antitops) are produced preferentially in the incoming quark (antiquark) direction. At hadron colliders it is difficult to determine the quark/antiquark direction, so another quantity in the laboratory frame is needed to measure this asymmetry. The variable $\Delta y= y_{t} - y_{\bar t}$, where $y$ represents the rapidity of the top/antitop quark, measured in the laboratory frame, is Lorentz invariant and it has the same value as the forward-backward asymmetry in the ${t\bar t}$ center-of-mass frame, computed as a function of the $\cos\theta^*$ angle between the top and the incoming quark. $\Delta y$ is the variable chosen by TeVatron experiments to measure this asymmetry, counting the number of events where $\Delta y$ is positive or negative.\\
At the LHC, due to the symmetry of the incoming beams, an aymmetry based on the $\Delta y$ variable would vanish. Hence the variable in Eq.1 
\begin{equation}
\Delta |y|= |y_{t}| - |y_{\bar t}|
\end{equation}
has been chosen, based on the fact that quarks are more boosted than antiquarks, due to the different mean momenta carried by valence quarks and sea antiquarks. The asymmetry $A_C$ obtained counting the number of events where $\Delta |y|$  is positive or negative, is called top quark charge asymmetry.\\
%In the following, the measurements from the ATLAS and CMS experiments, classified into single lepton and dilepton channel, depending on the top quark decay channel, will be shown respectively in Section 3 and Section 4, before to give a summary in Section 5.
\section{Measurements in the single lepton channel}
\subsection{ATLAS results}
\begin{figure}[!htb]
\begin{center}
\epsfig{file=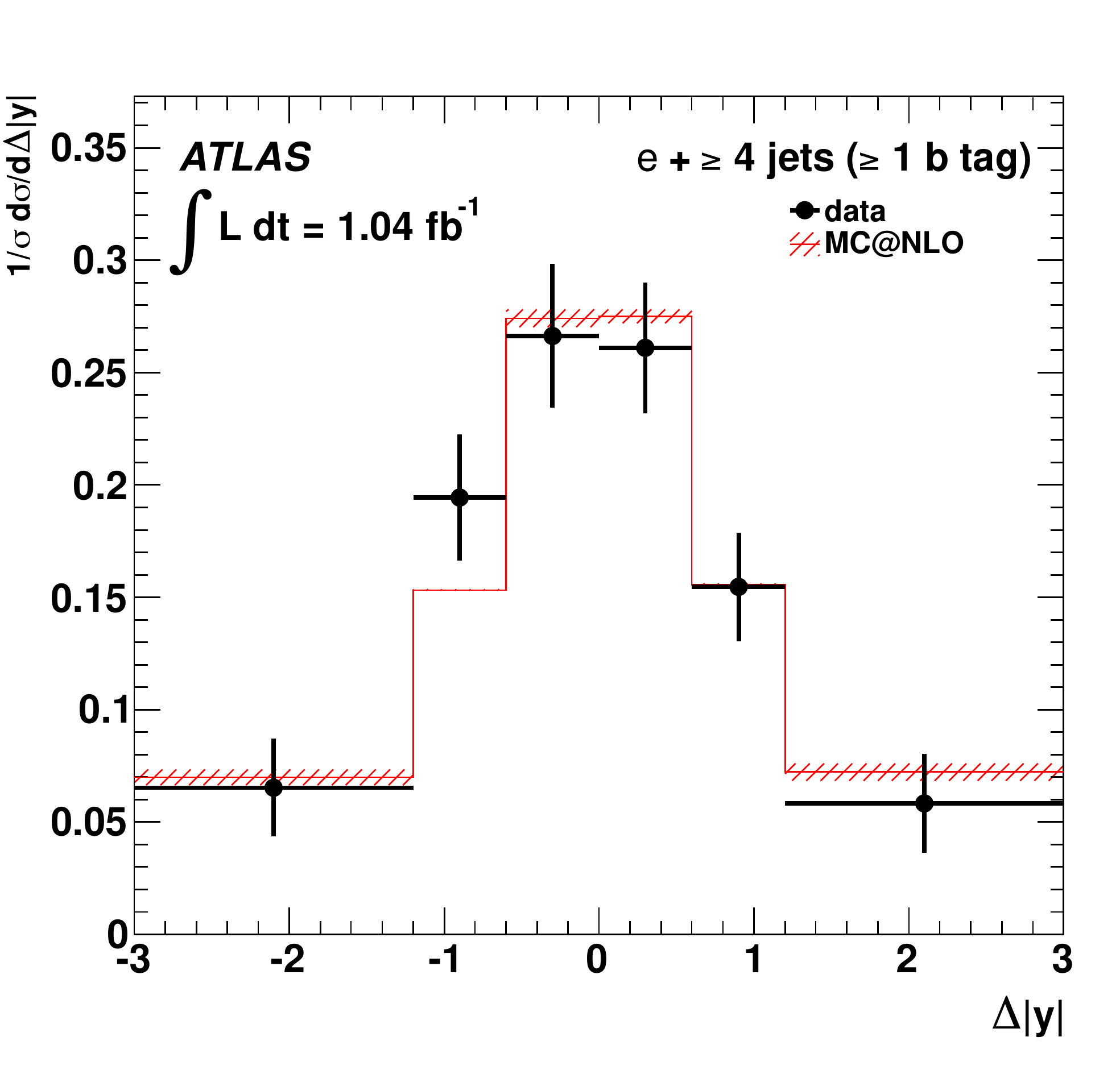,height=1.4in}
\epsfig{file=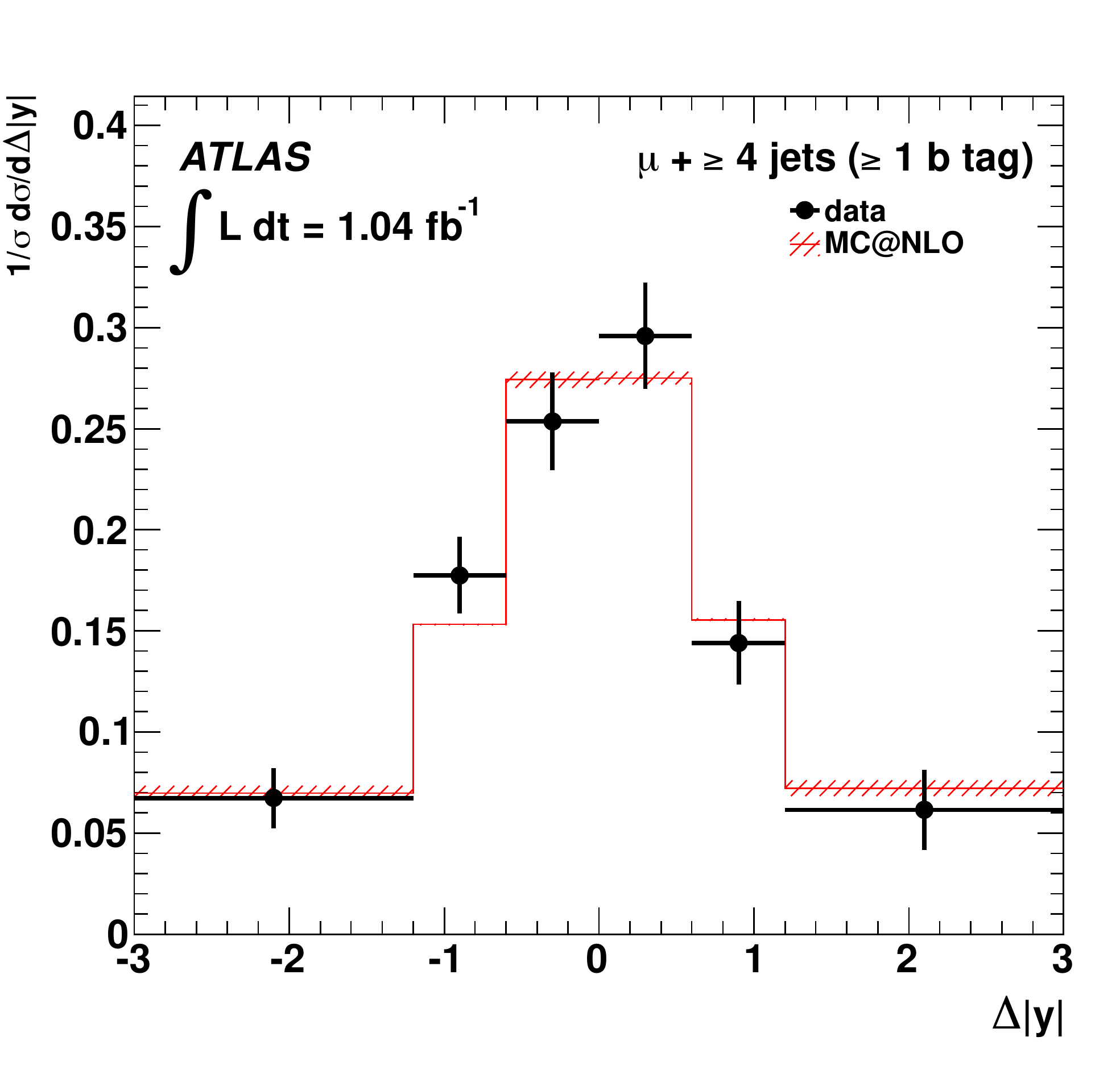,height=1.4in}
\epsfig{file=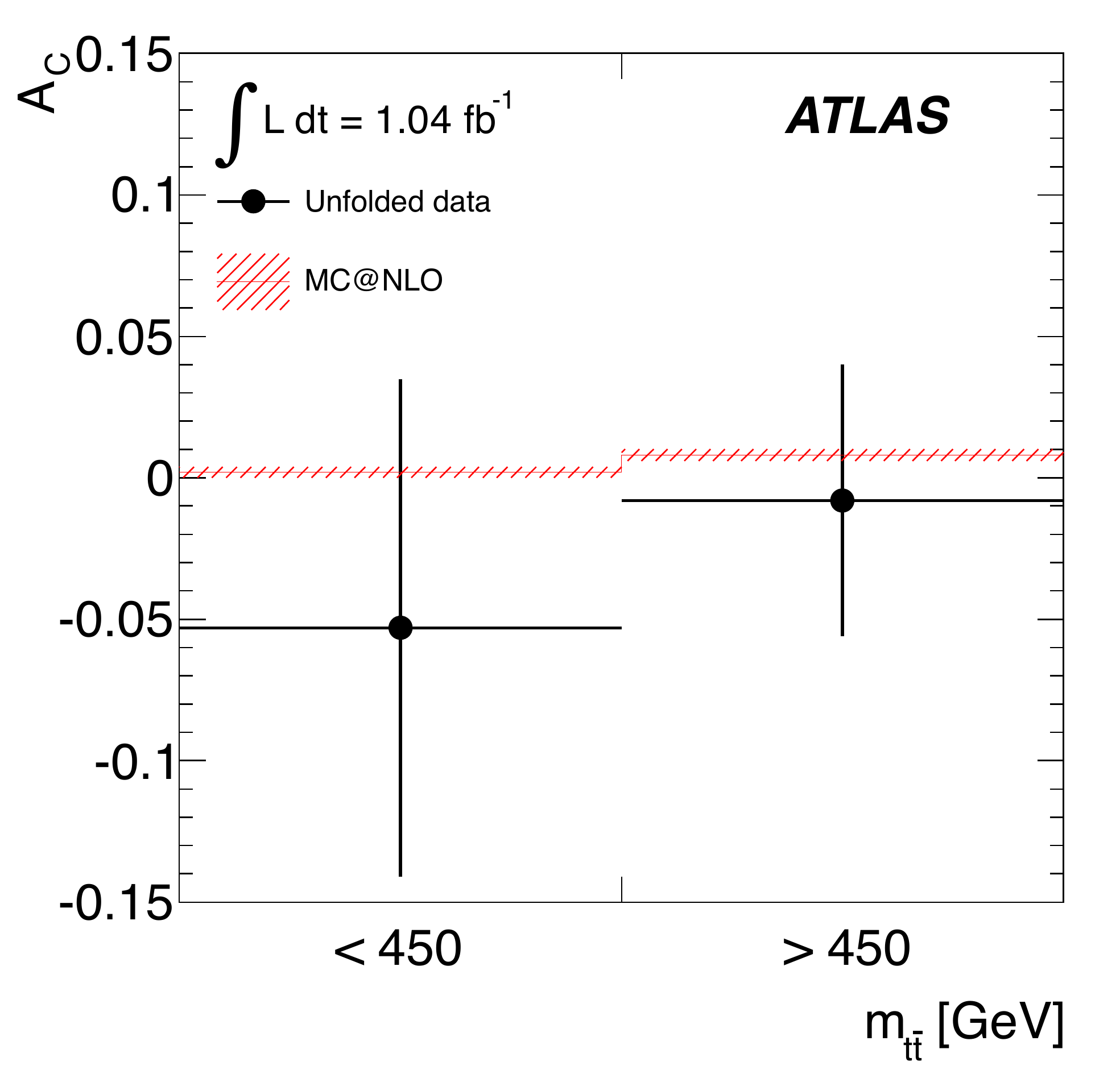,height=1.3in}
\caption{Inclusive unfolded $\Delta |y|$ distributions for electron (left plot) and muon channels (center plot) and the corrected asymmetry as a function of the $m_{t\bar{t}}$ (right plot) are shown. Each asymmetry is compared to the SM predictions obtained using the MC@NLO Monte Carlo generator. The uncertainties on the measurements include both statistical and systematic contributions~\cite{ATLAS}.}% The error bands on the MC@NLO prediction include uncertainties from parton distribution functions and renormalisation and factorisation scales.}
\label{fig:atlas1}
\end{center}
\end{figure}

The top quark charge asymmetry $A_C$ has been measured by the ATLAS experiment with data collected at 7 TeV center-of-mass energy corresponding to an integrated luminosity of 1.04 $\rm{fb}^{-1}$~\cite{ATLAS}. Events are selected requiring the presence of exactly one reconstructed isolated electron (muon) with $p_T>25$ (20) GeV, at least four jets (reconstructed with the anti-$k_T$ algorithm with a 0.4 radius parameter in the $\eta-\phi$ plane) with $p_T>25$~GeV of which at least one tagged as a $b$-jet. Additional cuts are applied on the missing transverse momentum $E_T^{miss} > 30$~GeV and the transverse W mass $m_T^W > 30$ GeV in the electron channel, while a cut on their sum $(E_T^{miss} + m_T^W) > 60$~GeV is applied on the muon channel. In both cases the aim is to reduce the multijet background.\\
The main backgrounds for this analysis, that are multijets and $W$+jets, are estimated using data-driven techniques, while minor backgrounds like $Z$+jets, single top and diboson ($WW$, $WZ$, $ZZ$) productions, are estimated using Monte Carlo simulated samples.\\
The $t\bar{t}$ system is then reconstructed using a kinematic likelihood fit that assesses the compatibility of the observed events with the topology of a simulated $t\bar{t}$ decays. This method identifies the correct decay topology in 74\% of the cases.\\
An unfolding procedure has been used in order to pass from the reconstructed asymmetries and  $\Delta |y|$ distributions to the partonic relative quantities. Detector effects as well as acceptance and efficiency induced effects are evaluated in simulated  $t\bar{t}$ events and used to build a response matrix relating true and reconstructed observed quantities. After the subtraction of the backgrounds to the data in the signal region, this matrix is inverted using a bayesian iterative procedure.\\
The inclusive charge asymmetry and the differential charge asymmetry as a function of the invariant mass of the $t\bar{t}$ pair ($m_{t\bar{t}}$) are measured separately in the electron and muon channels and then combined taking into account the correlations between the two measurements.
In Figure~\ref{fig:atlas1} the $\Delta|y|$ distributions after the unfolding procedure in both electron and muon channels are shown as well as the combined differential asymmetry as a function of $m_{t\bar{t}}$.\\
The inclusive charge asymmetry has been measured to be $A_C=-0.019 \pm 0.028 (\rm{stat.}) \pm 0.024 (\rm{syst.})$, while the differential asymmetries to be: $A_C=-0.052 \pm 0.070 (\rm{stat.}) \pm 0.054 (\rm{syst.})$  for $m_{t\bar{t}} < 450$ GeV and $A_C=-0.008 \pm 0.035 (\rm{stat.}) \pm 0.032 (\rm{syst.})$ for $m_{t\bar{t}} > 450$ GeV.\\

\subsection{CMS results}
\begin{figure}[htb]
\begin{center}
\epsfig{file=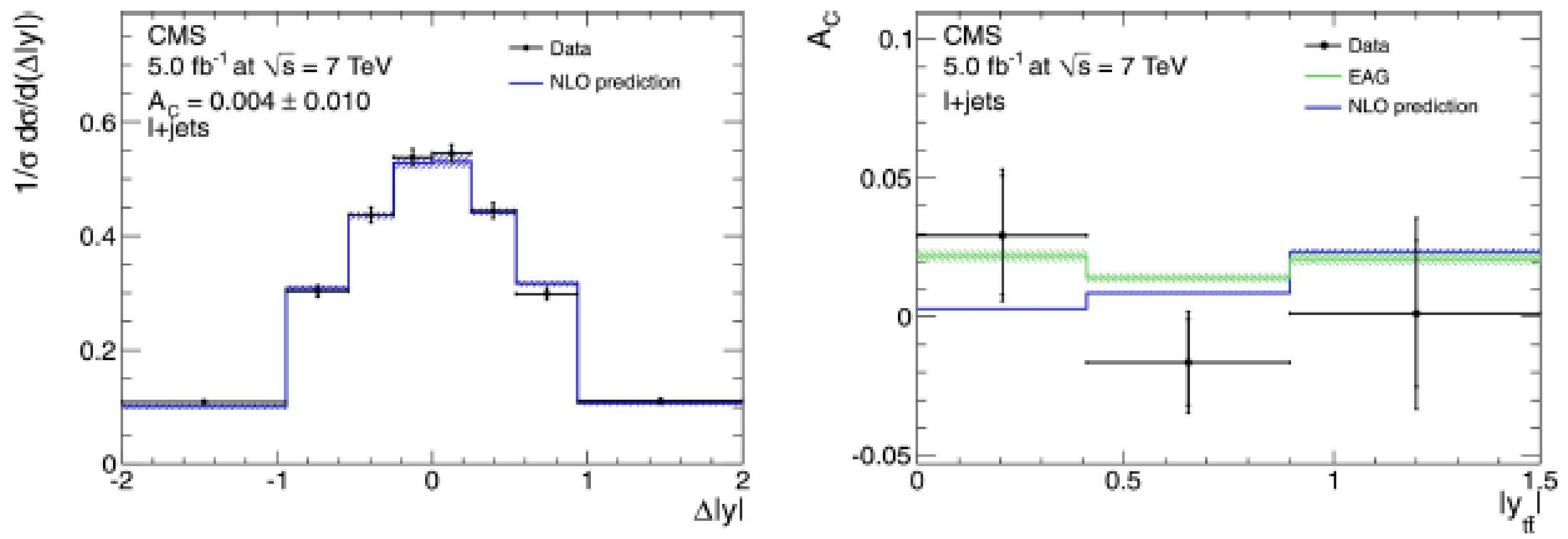,height=1.0 in}
\epsfig{file=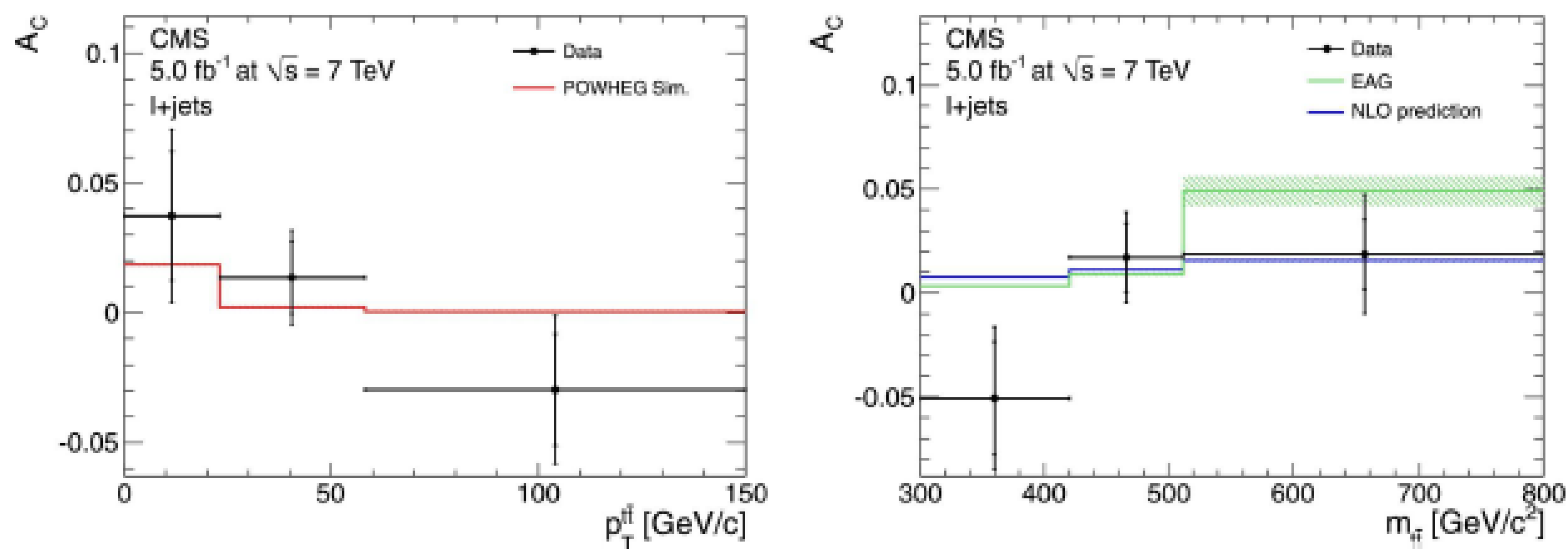,height=1.0 in}
\caption{From the left to the right: the unfolded inclusive $\Delta |y|$ distribution, the corrected asymmetry as a function of the rapidity $y_{t\bar{t}}$, the transverse momentum $p^T_{t\bar{t}}$ and the mass $m_{t\bar{t}}$ of the $t\bar{t}$ system compared to the SM predictions. The measured values are compared to NLO calculations for the SM and to predictions of an effective field theory (EFT). The error bars on the differential asymmetry values indicate the statistical and systematic uncertainties~\cite{CMS}.}
\label{fig:cms}
\end{center}
\end{figure}

The top quark charge asymmetry $A_C$ has been also measured by the CMS experiment with data collected at 7 TeV center-of-mass energy corresponding to an integrated luminosity of 4.7 $\rm{fb}^{-1}$~\cite{CMS}. Events are selected by an inclusive lepton trigger with the request to have exactly one reconstructed isolated electron (muon) with $p_T>30$ (20) GeV, at least four jets (reconstructed with the anti-$k_T$ algorithm with a 0.5 radius parameter in the $\eta-\phi$ plane) with $p_T>30$ GeV of which at least one tagged as a $b$-jet.\\
Multijets background is estimated using a specific data-driven technique, while the remaining backgrounds ($W$+jets, $Z$+jets, single top) are estimated using a binned likelihood fit. The background templates are taken from Monte Carlo simulated samples and the normalisation to data is extracted fitting the missing transverse momentum and the M3 (the invariant mass of the highest $p_T$ three jets combination) distributions. The $t\bar{t}$ system is reconstructed using a kinematic likelihood fit. Several jet-quark assignment hypotheses are tested and the one with the highest probability is taken. This method identifies the correct decay topology in 72\% of the cases.\\
An unfolding procedure has been used in order to pass from the reconstructed asymmetries and  $\Delta |y|$ distributions to the partonic relative quantities. To correct for these effects, a regularized unfolding procedure is applied. The unfolding algorithm corrects the measured spectrum by applying a generalized matrix inversion method. To regularize the problem and to avoid unphysical fluctuations two additional terms, a regularization term and a normalization term, are used.\\
The inclusive charge asymmetry and the differential charge asymmetry as a function of the invariant mass, the rapidity and the transverse momentum of the $t\bar{t}$ pair ($m_{t\bar{t}}$, $y_{t\bar{t}}$ and $p^T_{t\bar{t}}$ respectively) are measured combining the electron and muon channels. In Figure~\ref{fig:cms} the inclusive $\Delta|y|$ distribution as well as the three differential distributions mentioned above after the unfolding procedure are shown.\\
The inclusive charge asymmetry has been measured to be $A_C=0.004 \pm 0.010 (\rm{stat.}) \pm 0.012 (\rm{syst.})$.

\section{Measurements in the dileptonic channel}
A measurement of the charge asymmetry has been done by the ATLAS experiment also in the dileptonic $t\bar{t}$ decay channel with an integrated luminosity of 4.7~$\rm{fb}^{-1}$~\cite{ATLAS2}. The events are selected requiring exactly two oppositely charged leptons with the same flavor (i.e. $ee$, $e\mu$ and $\mu\mu$) and $p_T>30(20)$~GeV for electrons (muons). At least two jets with  $p_T>25$~GeV are also required. In $ee$ and $\mu\mu$ channels additional cuts are applied on the missing transverse momentum $E_T^{miss} > 60$~GeV and the invariant mass of the lepton pair $|m(ll)-m(Z)| < 10$~GeV to remove $Z$+jets background. In the $e\mu$ channel a cut on $H_T$, that is the scalar sum of the lepton and jets transverse momentum, is also applied: $H_T>130$~GeV.\\
Having two neutrinos in the final state, the kinematics of the $t\bar{t}$ decays is under-constrained. Hence several combinations of the physical objects in the final state are admissible for each event. In each event, each solution has been weighted according to a likelihood estimator derived from matrix elements for the LO process $gg\to t\bar{t}$. The combination with the highest weight is finally chosen.\\
To measure the asymmetry at the partonic level, a calibration procedure is used in this measurement. After the subtraction of the background, the measured asymmetry is calibrated using calibration curves that relate reconstructed and true asymmetries. The curves have been derived injecting in the simulation true asymmetries from $-10\%$ to $10\%$ and verifying the corresponding reconstructed asymmetries. Raw (after the background subtraction) and calibrated asymmetries in the various channels are shown in Table~\ref{tab:dil}. A combination among the $ee$, $e\mu$ and $\mu\mu$ channels has been then performed taking into account the correlations among the measurements. The combined asymmetry has been measured to be $A_C=0.057 \pm 0.024 (\rm{stat.}) \pm 0.015 (\rm{syst.})$. Furthermore, a combination with a single lepton channel measurement described in Section~3 has been performed, giving a combined asymmetry $A_C=0.029 \pm 0.018 (\rm{stat.}) \pm 0.014 (\rm{syst.})$.

\begin{table}[!]
\begin{center}
\begin{tabular}{l|cc}  
Channel &  Raw asymmetry & Calibrated asymmetry\\
\hline
$ee$ & $0.051 \pm 0.045 (\rm{stat.})$ & $0.079 \pm 0.087 (\rm{stat.}) \pm 0.028 (\rm{syst.})$ \\
$e\mu$ & $0.037 \pm 0.014 (\rm{stat.})$  & $0.078 \pm 0.029 (\rm{stat.}) \pm 0.017 (\rm{syst.})$ \\
$\mu\mu$ & $-0.001 \pm 0.022 (\rm{stat.})$ & $0.000 \pm 0.046 (\rm{stat.}) \pm 0.021 (\rm{syst.})$\\
\end{tabular}
\caption{Raw and calibrated top charge asymmetries in the three dileptonic channels.}
\label{tab:dil}
\end{center}
\end{table}

\section{Summary}
The top quark charge asymmetry measurements performed by ATLAS and CMS experiments have been presented. The asymmetry has been measured in the single lepton and dileptonic channel depending on the $t\bar{t}$ decay topology. Inclusive and differential measurements, as a function of specific kinematic quantities of the $t\bar{t}$ system has been also presented.
All the presented measurements did not show any significant deviation from the SM predictions.
%%%%%%%%%%%%%%%%%%%%%%%%%%%%%%%%%%%%%%%%%%%%%%%%%%%%%%%%%%%%%%%%%%%%%%%%%
%%
%%   use this format to include an .eps figure into your paper
%%
%\begin{figure}[htb]
%\begin{center}
%\epsfig{file=magnet.eps,height=1.5in}
%\caption{Plan of the magnet used in the Mesmeric studies.}
%\label{fig:magnet}
%\end{center}
%\end{figure}
%%%%%%%%%%%%%%%%%%%%%%%%%%%%%%%%%%%%%%%%%%%%%%%%%%%%%%%%%%%%%%%%%%%%%%%%%%%

%%%%%%%%%%%%%%%%%%%%%%%%%%%%%%%%%%%%%%%%%%%%%%%%%%%%%%%%%%%%%%%%%%%%%%%%%
%%
%%   use this format to include a LaTeX table  into your paper
%%

%%%%%%%%%%%%%%%%%%%%%%%%%%%%%%%%%%%%%%%%%%%%%%%%%%%%%%%%%%%%%%%%%%%%%%%%%%%

%\def\Discussion{
%\setlength{\parskip}{0.3cm}\setlength{\parindent}{0.0cm}
%     \bigskip\bigskip      {\Large {\bf Discussion}} \bigskip}
%\def\speaker#1{{\bf #1:}\ }
%\def\endDiscussion{}
%
%\Discussion
%
%\speaker{D. Giovanni (University of Seville)}  My analysis indicates that the
%recovery of the two gentlemen is due simply to their embrace of the masculine
%principle and has nothing to do with magnetism at all.  Could you comment on 
%this?
%
%\speaker{Reggiano} Professor Giovanni has discussed this hypothesis in several
%forums, but, I do not believe there is anything in print.  I understand that
%he is spending his time in other pursuits.
%
%\speaker{D. Anna (University of Seville)}  In fact, my colleague Giovanni 
%has expressed opposite opinions on this question at various times, depending
%on the audience.  All of these testosterone-based theories are, of course,
%nonsense.
%
%\endDiscussion
 
\end{document}

%% file: econfmacros.tex
\def\beq{\begin{equation}}
\def\eeq#1{\label{#1}\end{equation}}
\def\eeqn{\end{equation}}

\def\beqa{\begin{eqnarray}}
\def\eeqa#1{\label{#1}\end{eqnarray}}
\def\eeqan{\end{eqnarray}}

\let\bar=\overbar

\def\Dslash{\not{\hbox{\kern-4pt $D$}}}
\def\dslash{\not{\hbox{\kern-2pt $\del$}}}

\def\msb{{\bar{\ssstyle M \kern -1pt S}}}